\documentclass[10pt,a4paper]{article}

\usepackage{thumbpdf,lmodern}

\usepackage{framed}
\usepackage{graphicx}
\usepackage{subcaption}
\usepackage{amsmath}
\usepackage{algorithm}
\usepackage{algorithmicx,algpseudocode}
\usepackage{mathtools}

\newcommand{\scout}{{SCOUT }}
\newcommand*{\img}[1]{%
	\raisebox{-.3\baselineskip}{%
		\includegraphics[
		height=\baselineskip,
		width=\baselineskip,
		keepaspectratio,
		]{#1}%
	}%
}
\newcommand*{\imgg}[1]{%
	\raisebox{-.3\baselineskip}{%
		\includegraphics[
		height=\baselineskip,
		keepaspectratio,
		]{#1}%
	}%
}
\DeclarePairedDelimiter\floor{\lfloor}{\rfloor}

\author{Richard Semaan\\Technische Universit\"at Braunschweig
   \and Vikas Yadav\\Technische Universit\"at Braunschweig}
\title{{SCOUT}: Signal Correction and Uncertainty Quantification Toolbox in {MATLAB}}
\begin{document}
\maketitle
%% - \Abstract{} almost as usual
\begin{abstract}
	This manuscript describes the software package {SCOUT}, which analyzes, characterizes, and corrects one-dimensional signals. 
	Specifically, it allows to check and correct for stationarity, detect spurious samples, check for normality, check for periodicity, 
	filter, perform spectral analysis, determine the integral time scale, 
	and perform uncertainty analysis on individual and on propagated signals through a data reduction equation. 
	The novelty of \scout lies in combining these various methods into one compact and easy-to-use toolbox,
	which enables students and professionals alike to analyze, characterize, and correct for signals without expert knowledge.
	The program is oriented towards time traces, but an easy adaptation to spatial distributions can be performed by the user.
	\scout  is available in two variants: a graphical user interface (GUI) and a script-based version.
	A key motivation of having two variants is to offer maximum flexibility to adaptively and visually adjust the analysis settings using the GUI version  
	and to enable large batch processing capabilities and own code-integration using the script-based version. 
	The package includes both variants as well as three example scripts with their corresponding signals.
\end{abstract}

%% -- Introduction -------------------------------------------------------------

%% - In principle "as usual".
%% - But should typically have some discussion of both _software_ and _methods_.
%% - Use {}, {}, and \texttt{} markup throughout the manuscript.
%% - If such markup is in (sub)section titles, a plain text version has to be
%%   added as well.
%% - All software mentioned should be properly \cite-d.
%% - All abbreviations should be introduced.
%% - Unless the expansions of abbreviations are proper names (like "Journal
%%   of Statistical Software" above) they should be in sentence case (like
%%   "generalized linear models" below).
% =====================================================================
%                                                             SECTION
% =====================================================================
\section{Introduction} \label{sec:intro}

Signal processing and uncertainty quantification are two very broad yet related research areas. 
They have applications in many fields ranging from engineering to mathematics, to music.
With the proper tools, signal analysis allows us to discover valuable traits and characteristics in signals,
whereas uncertainties quantification informs us about the origin, propagation, and interplay of different sources of errors.
Due to their relevance, many open-source and commercial software packages have been developed,
such as the {SciPy}  \cite{Scipy2001} and the {Signal Processing Toolbox}  \cite{SigProcToolbox} packages for signal analysis, and {UQLab} \cite{MarelliICVRAM2014} and {propagate} \cite{Propagate2018} packages for uncertainty propagation.
These packages, and others, offer an impressive range of capabilities and options but are mainly geared toward expert users in either field.
This limits their usability in many engineering fields. 

In this manuscript, we present {SCOUT}, an easy-to-use signal processing and uncertainty quantification {MATLAB} package that is well suited to today's students and professionals alike.
It offers the main tools necessary to analyze, categorize, and quantify the uncertainty of acquired signals.
Specifically, it allows to check and correct for stationarity, detect spurious samples, check for normality, check for periodicity, filter,  
perform spectral analysis, determine the integral time scale, and perform uncertainty analysis on individual and on propagated signals through a data reduction equation. 
The uncertainty analysis yields uncertainties on central moment up to the fourth moment.
\scout  is available in two variants: a graphical user interface (GUI) version we label {SCOUT GUI}, and a script-based version we call {SCOUT Script}.
The GUI version offers maximum flexibility to adaptively and visually adjust the analysis settings, 
whereas the script version enables large batch processing capabilities and own code-integration.

\scout  is available as free software and can be downloaded from \\
https://www.richardsemaan.com/home/software/, with all supplementary materials which include both variants and three example scripts with their corresponding data.

%This manuscript has three main sections. 
%Section~\ref{sec:characteristics} presents general characteristics common to both of \scout's flavors.
%Section~\ref{sec:Interface} explains the .  
%After reading this chapter, the user should understand the various input variables and be able to conduct the complete analysis. 
%Section~\ref{chap:Script} describes the script-based version and how to run it.

%% -- Manuscript ---------------------------------------------------------------

%% - In principle "as usual" again.
%% - When using equations (e.g., {equation}, {eqnarray}, {align}, etc.
%%   avoid empty lines before and after the equation (which would signal a new
%%   paragraph.
%% - When describing longer chunks of code that are _not_ meant for execution
%%   (e.g., a function synopsis or list of arguments), the environment {Code}
%%   is recommended. Alternatively, a plain {verbatim} can also be used.
%%   (For executed code see the next section.)

% =====================================================================
%                                                             SECTION
% =====================================================================
\section{General Characteristics} \label{sec:characteristics}

The two \scout versions, the graphical user interface (GUI) and the script versions are intended to be complimentary.
{SCOUT GUI} provides visual as well as numerical output at every step.
It allows dynamic and adaptive analysis with a range of options and parameter tuning. 
On the other hand, {SCOUT Script} is command-based designed for integrated and batch processing.
In other words, it allows all of {SCOUT}'s capabilities to be integrated within the user's own analysis code.
In addition, the script version enables batch processing of large data.
An envisioned workflow scenario would be to perform initial analysis using {SCOUT GUI}, where the settings are fine-tuned, 
followed by the {SCOUT Script} for integrated or batch processing.
In the following, we list general characteristics common to both versions:
\begin{itemize}
	\item Each signal to be imported should \emph{only} include the time trace without the time vector, i.e., a one-dimensional array. 
	\scout generates the time vector from the specified sampling frequency and the inferred record length. Including more than one signal in the file will result in only one signal imported; the first one that is read.
	\item All signals \emph{must} be sampled at a constant sampling frequency $f_s$ (i.e., equal time interval $\Delta t$ between the samples).
	Moreover, if uncertainty propagation is sought, all propagated signals should be sampled at (or decimated to) the same rate.
	\item Up to 5 signals can be simultaneously analyzed, and, if desired, their uncertainty propagated.
	\item It is recommended that the user performs the analysis in the  presented order, i.e., `Stationarity Analysis', followed by `Spurious Samples', and so on.
\end{itemize}

% =====================================================================
%                                                             SECTION
% =====================================================================
\section{Interface and Code Execution}\label{sec:Interface}

The two \scout versions have two very different interfaces.
While the GUI version is adaptive, the script version is static and requires a configuration file.
This section details both interfaces.

\subsection[SCOUT GUI Interface]{{SCOUT GUI} Interface}

\begin{figure}
	\centering
	\includegraphics[width=\linewidth]{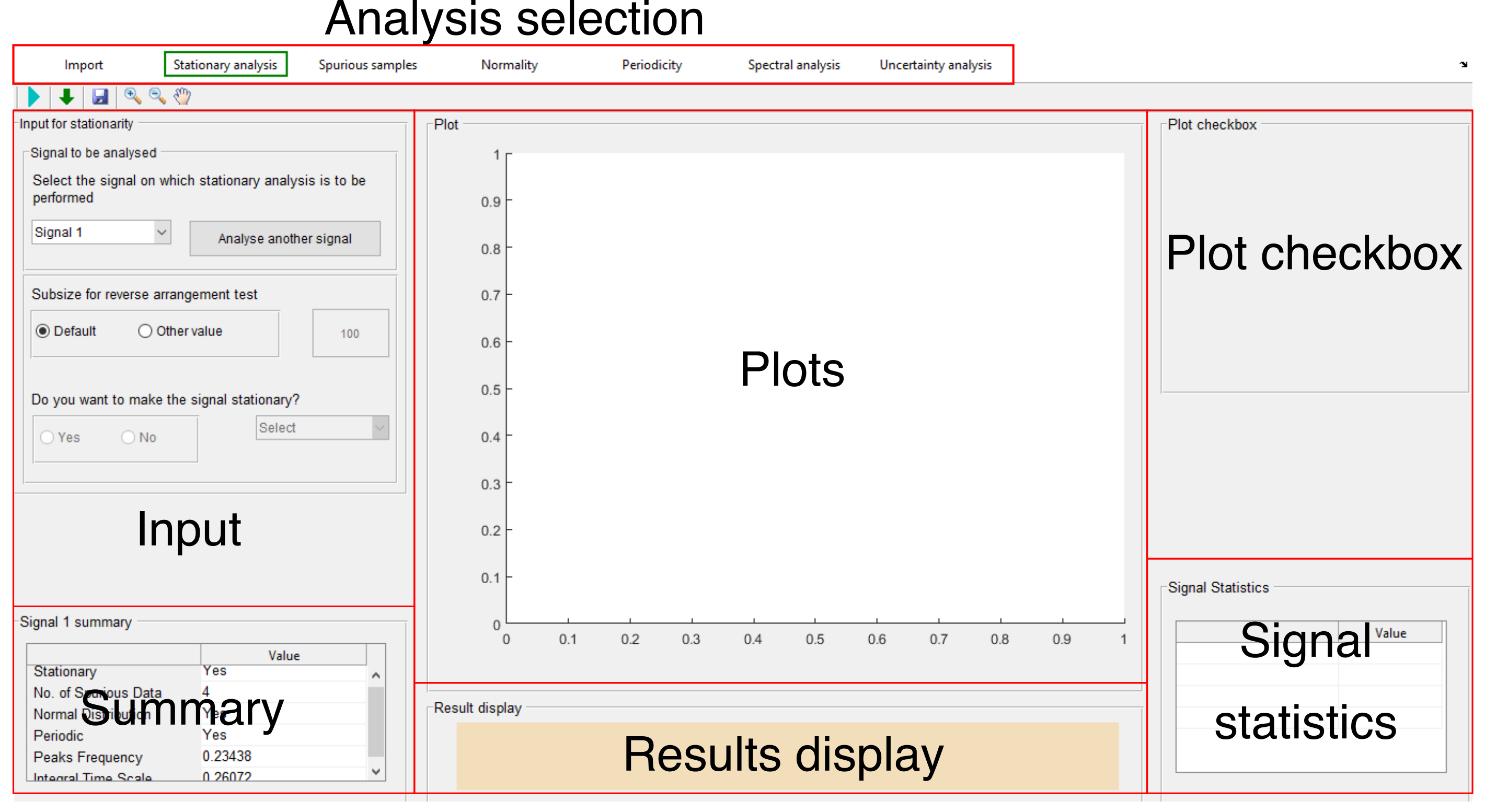}
	\caption{The general layout of {SCOUT GUI}. The red boxes highlight the various sections of the interface.}
	\label{fig:generallayout}
\end{figure}

A screenshot of {SCOUT GUI}'s interface is presented in Figure~\ref{fig:generallayout}.
The red boxes highlight the various regions of the layout, which are detailed below.
The layout structure is general and applies to most of the analysis steps.
It has the following general characteristics:
\begin{itemize}	
	\item The user is required to press the green \img{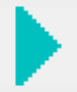} button to execute an analysis step.
	\item Hovering over the \img{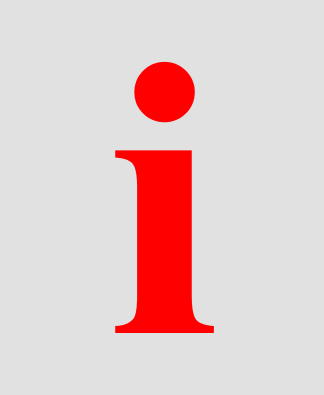} buttons provides information and tips.
	\item Each figure can be manipulated and saved using these buttons \imgg{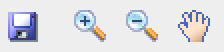} on the top left.
	\item The user can generate and download a summary file using the \img{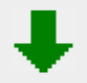} button.
	\item Each signal is constantly updated and logged in the workspace with each analysis step. This enables the user to select any previously-analyzed signal version for further examination.
\end{itemize}
% ============================== FIGURE===============================

For the remainder of this section, we detail the various interface sections.
\subsubsection{Analysis selection section}
The analysis selection section consists of seven tabs for the various analyses.
It is recommended the user performs the analyses in the suggested order from left to right.
In other words, after importing the signal(s), the next recommended analysis step is to check for stationarity, followed by checking for spurious samples, and so on.
It is worth to note that the user is not \emph{obligated} to perform all suggested steps.
Each processing step can be independently conducted.

\subsubsection{Input section}
The input section requires the user to input the information necessary for each process.
The input type ranges from numerical values to checkboxes, to drop-down menus.
The input section changes with each analysis tab.

\subsubsection{Summary section}
The summary section displays the latest general summary of the signal analyzed.
This section is constantly updated after each analysis step.
The summary section displays the following information about \emph{each} signal:
\begin{itemize}
	\item Signal stationarity status
	\item Number of spurious samples
	\item Whether the signal is normally distributed
	\item Whether a periodic component is removed
	\item The detected peak frequency
	\item The integral time scale
\end{itemize}

It is worth to note that this section distinguishes itself from the `Signal statistics' in that it does not display any statistical information about the signals.

\subsubsection{Plot and plot checkbox sections}

The plot section in the middle is reserved for plotting the various results. Plots range from time traces to histograms, to spectra. 
The user can toggle between various signals using the checkboxes on the right side.

\subsubsection{Results display}

The purpose of the results display section is to `announce' important results after the completion of a certain analysis step. This includes the following:

\begin{itemize}
	\item Whether the signal is stationary with 95\% confidence.
	\item Whether the signal is normally distributed with the specified level of confidence.
	\item The frequency of the highest detected peak in the spectrum.
\end{itemize}

\subsubsection{Signal statistics}
The signal statistics section displays the latest statistical information of the signal being analyzed. 
This section is constantly updated after each analysis step. The signal statistics section displays the following statistical information about \emph{each} signal:
\begin{itemize}
	\item Mean 
	\item Standard deviation
	\item z-score (for stationary analysis)
	\item Kurtosis 
	\item Skewness 
\end{itemize}

\subsection[GUI Script Interface]{{GUI Script} Interface}
{SCOUT Script} requires one post-processing configuration file for each imported signal, and a single optional uncertainty analysis configuration file.
The configuration files contain the various analysis settings.
Hence, unlike {SCOUT GUI}, all analysis settings are fixed during the execution.

The code execution is performed by calling on the main function \texttt{SCOUT\_Script} followed by the various configuration files:\\
$\texttt{ >>SCOUT\_Script(ConfigFile1(`U',`Workscape',`u',10000),UncertaintyConfigFile) }$
This code snippet, taken from \texttt{Example1.m} of the package, shows how \texttt{SCOUT\_Script} handles the two types of configuration files:
\begin{enumerate}
	\item Configuration file type 1 (e.g., \texttt{ConfigFile1}): is required for the entire analysis sequence \emph{except} for the uncertainty analysis (the last tab in {SCOUT GUI}).
	\item Configuration file type 2 (e.g., \texttt{UncertaintyConfigFile}): to specify the uncertainty analysis settings.
\end{enumerate}
In case uncertainty analysis is not sought, the user can simply withhold the type 2 configuration file.
Each type 1 configuration file requires four direct inputs:
\begin{enumerate}
	\item The name of the output summary file.
	\item The location of the signal (\texttt{Workspace} or \texttt{Directory}).
	\item The name of the signal in case the file is loaded from the workspace, or the name of the file including its path in case the signal is loaded from a directory.
	\item The sampling frequency.
\end{enumerate}
These input options allow batch processing of different signals with different names, saved at different locations, and sampled at different rates.
The reader is referred to \texttt{Example3} in the package for a batch processing example script.

Similar to the GUI version, the script version can simultaneously process five signals.
A code execution example for simultaneously analyzing two signals with two different configuration files without uncertainty analysis would be (refer to \texttt{Example2.m}): \\
$\texttt{ >>SCOUT\_Script(ConfigFile1(`U',`Workscape',`u',10000), ...}$\\
$\texttt{ConfigFile2(`V',`Directory',`Data/Sig2.mat',10000))}$

Besides the three direct input variables, the script version requires a list of other settings in the configuration files. 
The various analyses offered and their corresponding settings are detailed in section~\ref{sec:Analyses}.
The user is also referred to the example scripts in the package and the therein-included comments for further details.

It is worth to note that unlike {SCOUT GUI}, {SCOUT Script} does not offer the option to select different signal variants at different stages of the analysis. 
Specifically,  {SCOUT Script} sequentially performs all analysis steps activated in the configuration file always using the most up-to-date signal from the previous step.
For example, if the users wishes to analyze a signal \emph{without} removing any possible spurious samples, they are simply required to deactivate the relevant variables in the configuration file type 1 by setting: \texttt{I.Spurious.Chauvenet.n = `NA'} and \texttt{I.Spurious.Postprocessing = `NA'}. 
The variables in the aforementioned scenario are detailed in section~\ref{sec:Spurious}.

% =====================================================================
%                                                             SECTION
% =====================================================================
\section{Analyses}\label{sec:Analyses}

All analyses begin with importing signals. Up to 5 signals can be simultaneously imported. 
In {SCOUT GUI}, the total number of signals to be imported is provided as an option at the top of the input section. 
Each signal can be either imported from the drive or {MATALB}'s workspace. 
The user is required to provide the sampling frequency for each signal. 
It is important to note that when a file containing several signals, \emph{only} the first read signal will be imported. 
%For this reason, it is recommended to import files with only one signal per file. 

After importing the signal(s), the user is offered a range of analysis possibilities.
This section briefly presents the theory behind each analysis step and details the corresponding options.

% =====================================================================
%                                                             STATIONARITY
% =====================================================================
\subsection{Stationarity analysis}

This analysis examines and optionally renders signals stationary. 
The stationarity check is based on the reverse arrangement test \cite{SiegelCastellan88},
which we will briefly review.
For more details about the method, the reader is referred to the ample literature on the subject \cite{SiegelCastellan88,bendat_random_2011}.

The reverse arrangement test first requires the sample record to be specifically prepared.
The signal is split into $M$ equal time intervals, where the data in each interval may be considered independent.
For each interval, the mean square value is then computed as
\[
\overline{x_1^2}, \, \overline{x_2^2}, \, \overline{x_3^2}, \, \dotsc, \, \overline{x_M^2}\,.
\]
The reverse arrangement test for stationarity is then performed on the mean square values.
The process involves counting the number of times that $\overline{x_i^2} > \overline{x_j^2}$ for $i < j$. 
Each such inequality is called a reverse arrangement. The total
number of reverse arrangements is denoted by $A$,
where $A$ is defined from the set of observations $\overline{x_1^2}, \overline{x_2^2}, \dotsc, \overline{x_M^2}$ as follows,
\begin{equation*}
h_{ij}=\left\{
\begin{array}{ll}
1 \quad \text{if} \quad \overline{x_i^2}>\overline{x_j^2}\\
0     \quad \text{if} \quad \text{otherwise}
\end{array}
\right.
\end{equation*}
Then
\[ 
A=\sum_{i=1}^{M-1}A_i\,,
\]
where
\[
A_i=\sum_{j=i+1}^{M}h_{ij}\,.
\]
$A$ is then compared to the $z$-score, which is defined as,
\begin{equation*}\label{eq:zScore}
z=\frac{A-\left[ \frac{M(M-1)}{4} \right] }{\sqrt{\frac{2M^3+3M^2-5M}{72}}}\,.
\end{equation*}
The null hypothesis for this test is that the data points in the signal are independent
observations from a stationary random variable. 
To reject the null hypothesis at 95\% confidence level requires a $z$-score of
$z \geq 1.96$ or $z  \leq -1.96$. 

Stationary analysis requires only one input; The sample size for the reverse arrangement test, which should be a positive number \texttt{1 < M < length(signal)}. 
The default sample size is 100. The sample size should be chosen such that it is neither too short nor too long with respect to the fundamental period of the signal.

If the signal is not stationary, the options for rendering it stationary become active.  Signals can be made stationary using two methods:
the {MATLAB} inbuilt \texttt{detrend} function, or the proposed \texttt{Polynomial fit} method.
{MATLAB}'s \texttt{detrend} function simply detects and removes linear trends in the data. 
The user is referred to {MATLAB} user manual for details.
The polynomial fit method uses sequentially higher polynomial orders up to third order\footnote{Higher order polynomials than the third are not recommended \cite{bendat_random_2011}.} to fit the data, and then automatically chooses the best polynomial order based on a compromise between accuracy (fit error) and complexity (polynomial order). 
The steps are summarized in Algorithm~\ref{alg:PolOrder}.
\begin{algorithm}
	\caption{ Polynomial order identification for stationarity analysis}\label{alg:PolOrder}
	\textbf{Input:} non-stationary signal
	\begin{algorithmic}[1]
		\State Fit three polynomial curves with order 1--3 to the signal
		\If{$E_1< E_2, E_3$}             \Comment{$E_i$ is the fit standard error}
		\State  Select $\mathcal{O}=1$ \Comment{$\mathcal{O}$ is the polynomial order}
		\Else 
		\State Fit a spline curve through $E_i/E_1$ and $\mathcal{O}_i$, for $i=1,2,3$ 
		\State Compute the gradient of the spline curve $\mathcal{G}(E)$
		\State Infer $\mathcal{O}_{min}=\min(\mathcal{G}(E))$
		\State $\mathcal{O}= \floor{\mathcal{O}_{min}}-0.2$  \Comment{$\floor{\> }$ denotes the floor rounding function}
		\EndIf\\
		\Return Polynomial order
	\end{algorithmic}
\end{algorithm}

\begin{figure}
	\begin{subfigure}[b]{0.45\textwidth}
		\includegraphics[width=\textwidth]{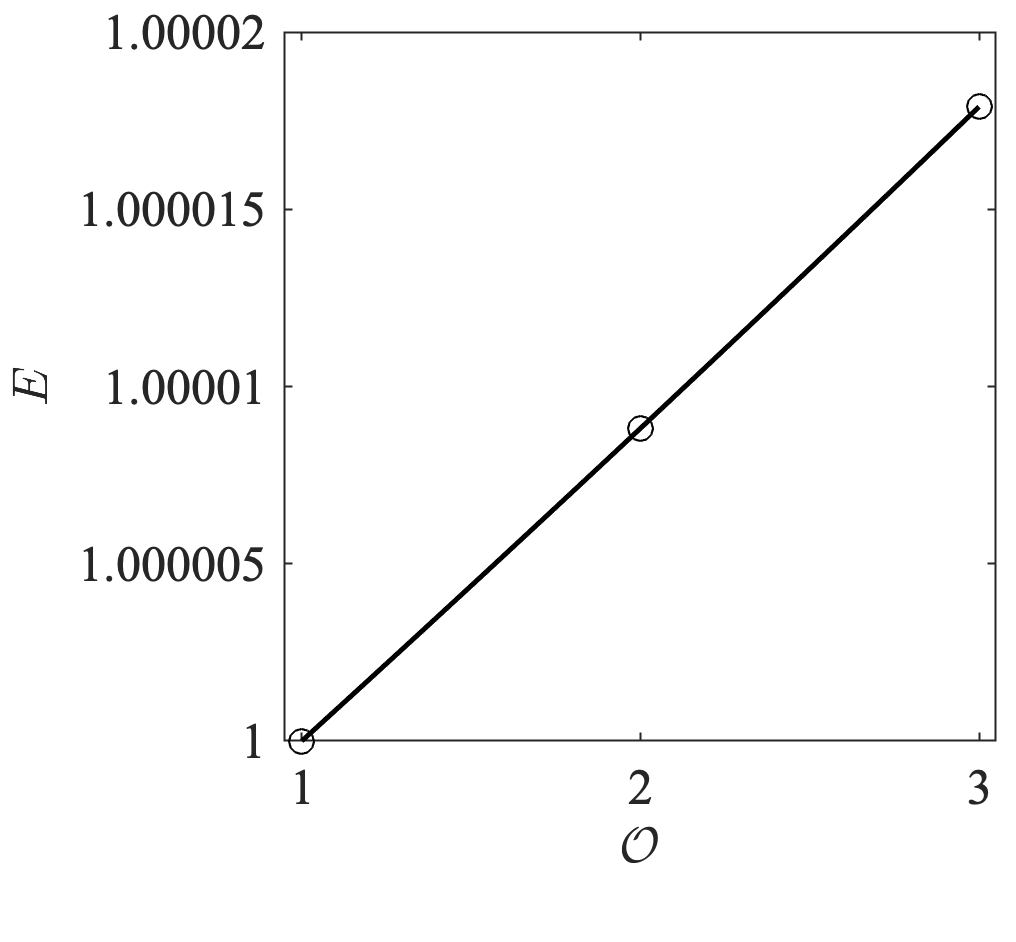}
		\caption{Selected polynomial order: $\mathcal{O}=1$.}
		\label{fig:f1}
	\end{subfigure}
	\hfill
	\begin{subfigure}[b]{0.45\textwidth}
		\includegraphics[width=\textwidth]{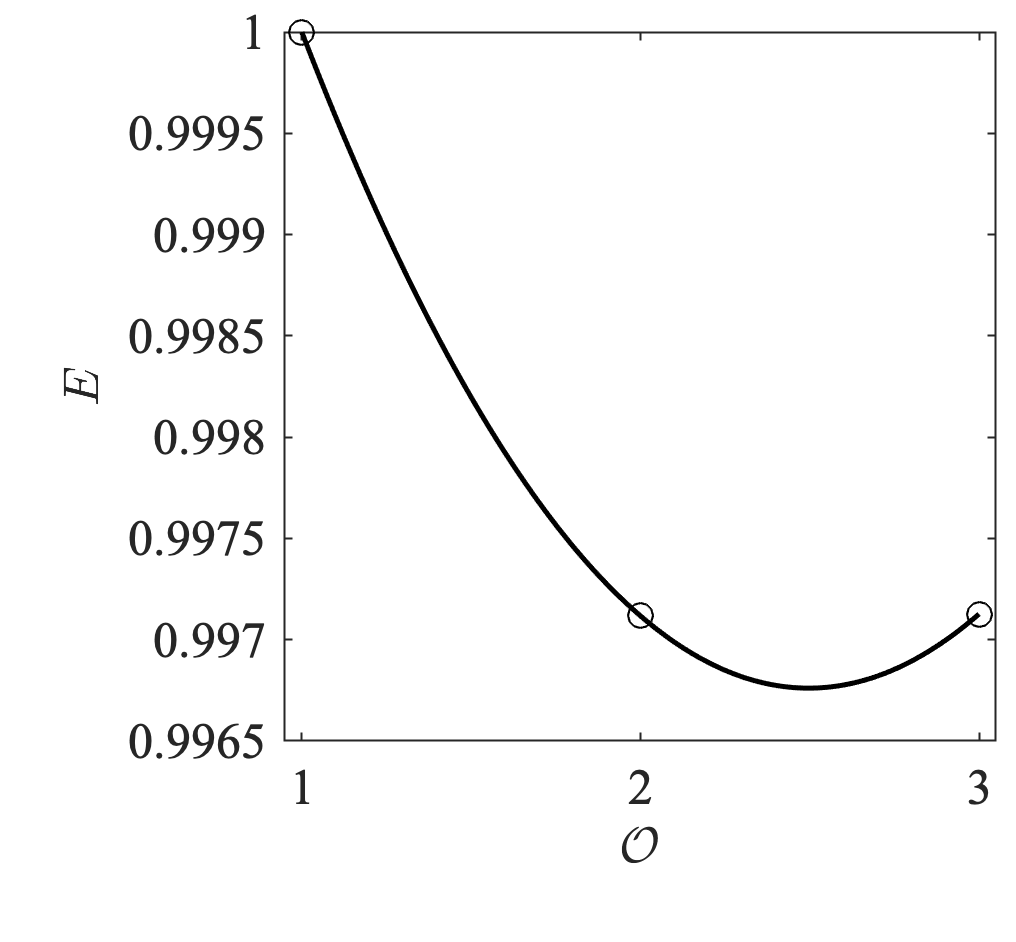}
		\caption{Selected polynomial order: $\mathcal{O}=2$.}
		\label{fig:f2}
	\end{subfigure}
	\caption{The polynomial order selection for detrending based on the error distribution fit. Two scenarios showing the selection outcome for (a) a first order $\mathcal{O}=1$ and for (b) a second order $\mathcal{O}=2$ polynomial.} \label{fig:PolySelection}
\end{figure}

In the last step of Algorithm~\ref{alg:PolOrder}, an explicit bias of -0.2 is introduced to favor simplicity.
The algorithm is tested on two non-stationary signals of polynomial order 1 and 2 with large superimposed noise.
As Figure~\ref{fig:PolySelection} shows, the resulting polynomial orders are correctly identified.
After the detrending process, the reverse arrangement test is again repeated to verify stationarity. 

In {SCOUT GUI}, the process can be separately performed on all imported signals, 
with every signal individually selected from the drop-down menu. 
Each time a new signal or a new detrending method is selected the user must press the \img{Figures/Play} button to execute the analysis. 
When a new signal is selected, all settings return to default.
The main steps of the detrending process are registered and updated in the `Signal summary' section.
Similarly, the `Signal statistics' section gets updated with the new statistical values from the detrended signal.

In {SCOUT Script}, the stationarity process is controlled through 3 variables in the configuration file:
\begin{itemize}
	\item \texttt{I.Stationarity.Subsize}: the sample size for the reverse arrangement test.
	\item \texttt{I.Stationarity.MakeStationary}: a binary selection (\texttt{Yes/No}) on whether to render the signal stationary.
	\item \texttt{I.Stationarity.StationaryMethod}: the choice of the detrending method.
\end{itemize}

% ============================== FIGURE===============================

% =====================================================================
%                                                             SPURIOUS SAMPLES
% =====================================================================
\subsection{Spurious Samples}\label{sec:Spurious}

\begin{figure}
	\centering
	\includegraphics[width=0.8\linewidth]{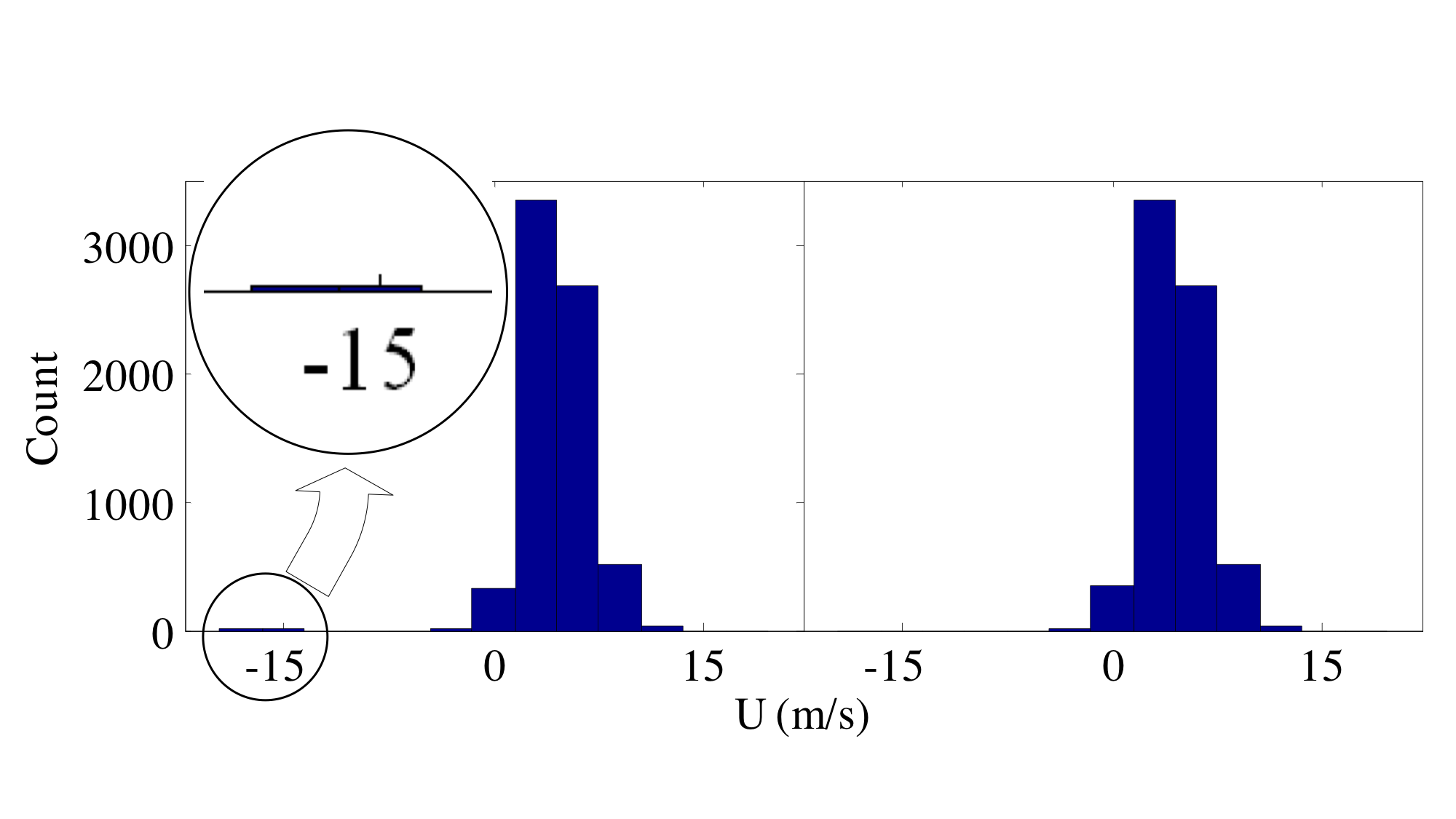}
	\caption{Illustration of the detection and the elimination of spurious points using the histogram method \cite{semaan2013}.}
	\label{fig:SpuriousDetection}
\end{figure}

This section describes the detection method and the handling of spurious samples.
Two possible algorithms for detecting spurious samples are offered: the Chauvenent criterion, and the histogram method.
The Chauvenet criterion \cite{chavuenet1960, coleman_experimentation_2009} is a widely accepted method for spurious sample detection.
It specifies that all points that fall within $1-1/(2N)$ probability band around the mean value should be retained,
where $N$ is the sample record length.
The Chauvenet criterion can be used on most random data, except when the measured probability density function is skewed or multi-modal, which causes a  rejection bias, where `real’ data get clipped. 
Under this scenario, the so-called histogram approach should be used instead.
The histogram approach \cite{semaan2013} simply constructs a coarse histogram and detects outliers as samples that are separate from the main histogram body.
This approach is illustrated in figure \ref{fig:SpuriousDetection}, where the spurious point in the original data on the left side has been removed, as shown on the right side.
% ============================== FIGURE===============================

If spurious samples are detected, the options to deal with them become active. 
In this case, the user has two options: removal, or replacement.
As the name suggests, the removal option simply deletes the spurious samples.
Removal is the preferred option when no subsequent spectral analysis is planned since any Fourier transform requires data sampled at equal time intervals.
Specifically, this will exclude performing `Periodicity', `Spectral analysis' and `Uncertainty analysis'.
If any of the above-mentioned analysis is desired, the replacement option should be selected. 
Here, spurious samples are replaced by their local average values.

In {SCOUT GUI}, the main steps of the spurious sample detection are visualized in red in the time trace and the histogram plots, and are reported and updated in the `Signal summary' and the `Signal statistics' sections.

In {SCOUT Script}, this analysis is controlled with 4 input variables in the configuration file:
\begin{itemize}
	\item \texttt{I.Spurios.DetectionMethod}: the choice of the detection method.
	\item \texttt{I.Spurious.Chauvenet.n}: the extent of the sample retention region away from the mean as a number of standard deviations. The default value is the one computed from the Chauvenet criteria.
	\item \texttt{I.Spurious.Histogram.Bins}: the number of bins for the histogram method.
	\item \texttt{I.Spurious.Postprocessing}: selection between the removal and the replacement options.
\end{itemize}

% =====================================================================
%                                                             NORMALITY
% =====================================================================
\subsection{Normality}
This section describes the method to identify normally-distributed signals to a pre-selected confidence level.
Among many motivations, knowing whether a signal is Gaussian yields significant simplifications in the uncertainty analysis.
The method to check for normality is the $\chi^2$ goodness-of-fit test \cite{cochran1952}, which is used in a wide range of applications.
In the following, we shall briefly describe it.

Consider a sample record of $N$ independent observations {\footnote{\scout offers the possibility to compute the integral time scale, and hence to identify the independent samples. However, such capabilities are not integrated in the normality test. It is the user's responsibility to make sure the observations are independent. }} from a
random variable $x$ with a probability density function $p(x)$. Let the $N$ observations
be grouped into $K$ bins, which together form a frequency histogram. 
The number of observations falling within the $i$th bin is denoted by $f_i$. The number of
observations that would be expected to fall within the $i$th bin for the true normally-distributed
probability density function $p(x_0)$ are denoted by $F_i$. 
The discrepancy between the observed
frequency and the expected frequency within each bin is given by $f_i-F_i$. 
The measure of the total discrepancy for all bins is the sum of the normalized squares of the discrepancies in
each bin,
\begin{equation}\label{eq:X2Dist}
X^2=\sum_{i=1}^{K}\frac{(f_i-F_i)^2}{F_i}\,.
\end{equation}

The $X^2$ distribution in equation \eqref{eq:X2Dist} is similar to that of the $\chi^2_n$. 
To compare against $\chi^2_n$, one needs to know the degrees of freedom $n$,
which for a normal distribution is $n=K-3$.
The hypothesis is accepted if $X^2\leq \chi^2_{n;\alpha}$, where $\alpha$ is the level of significance.
The method is summarized in Algorithm~\ref{alg:chi2}.
\begin{algorithm}
	\caption{ $\chi^2$ goodness-of-fit test}\label{alg:chi2}
	\textbf{Input:} confidence level, number of bins
	\begin{algorithmic}[1]
		\State Group the sampled observations into $K$ bins  %\Comment{\textcolor{blue}{Recommended bin size is $\Delta x \approx 0.4\>s$, where $s$ is the data standard deviation.}}
		\State Compute the expected frequency for each interval $f_i$  %\Comment{$p_0(x)$ is the assumed normal distribution.}
		\State Compute $X^2$ as in equation \eqref{eq:X2Dist}
		\If{$X^2 > \chi_{n;\alpha}^2$}
		\State The hypothesis that $p(x)=p(x_0)$ is rejected at the $\alpha$ level of significance
		\Else 
		\State The hypothesis is accepted at the $\alpha$ level of significance
		\EndIf \\
		\Return Test result
	\end{algorithmic}
\end{algorithm}

The outcome of Algorithm~\ref{alg:chi2} is the test result on whether the distribution is normal.
In {SCOUT GUI}, the normality test result is displayed in the `Result display' section for the selected confidence level.
As a visual guide, a  normally-distributed reference probability density function $p(x_0)$ with the same mean and standard deviation as the signal is displayed in red in the histogram plot.
As before, the main steps of the normality test are registered and updated in the `Signal summary' section.

In {SCOUT Script}, the normality test is controlled via 2 variables in the configuration file:
\begin{itemize}
	\item \texttt{I.Normality.ConfidenceLevel}: the desired confidence level (to remain consistent and avoid confusion, the confidence level is used instead of the level of significance employed in Algorithm˜\ref{alg:chi2}).
	\item \texttt{I.Normality.NumberOfBins}: the number of bins. The default recommended bin size is $\Delta x \approx 0.4\>s$, where $s$ is the standard deviation.
\end{itemize}

% =====================================================================
%                                                             PERIODICITY
% =====================================================================
\subsection{Periodicity}\label{sec:Periodicity}
This section describes methods to detect periodic components and to optionally filter them.
Removing deterministic (e.g. periodic) components from signals is a necessary prerequisite before uncertainty quantification. 
The periodicity analysis starts with computing the autospectral density function \cite{welch1967}, which is a standard function in {MATLAB}.
The magnitude of the highest spectral peak, as well as its frequency, are then detected.

\begin{figure}
	\begin{subfigure}[b]{0.45\textwidth}
		\includegraphics[width=\textwidth]{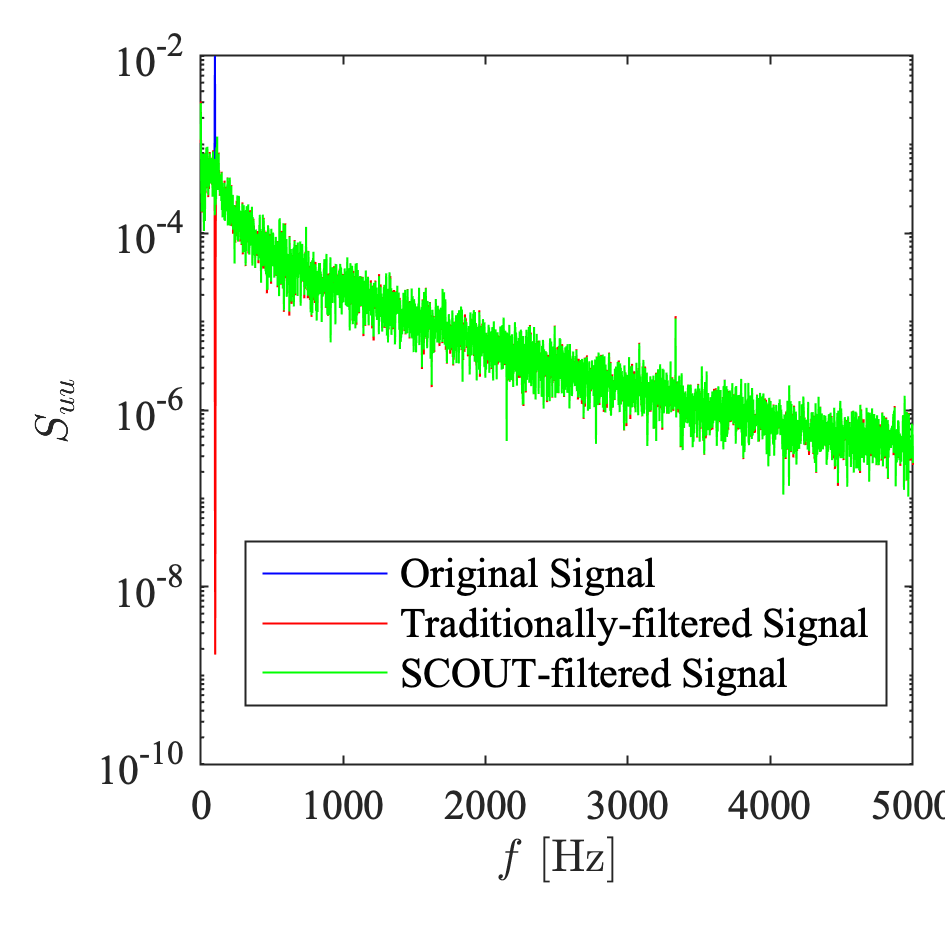}
		\caption{Entire frequency range.}
		\label{fig:f31}
	\end{subfigure}
	\hfill
	\begin{subfigure}[b]{0.45\textwidth}
		\includegraphics[width=\textwidth]{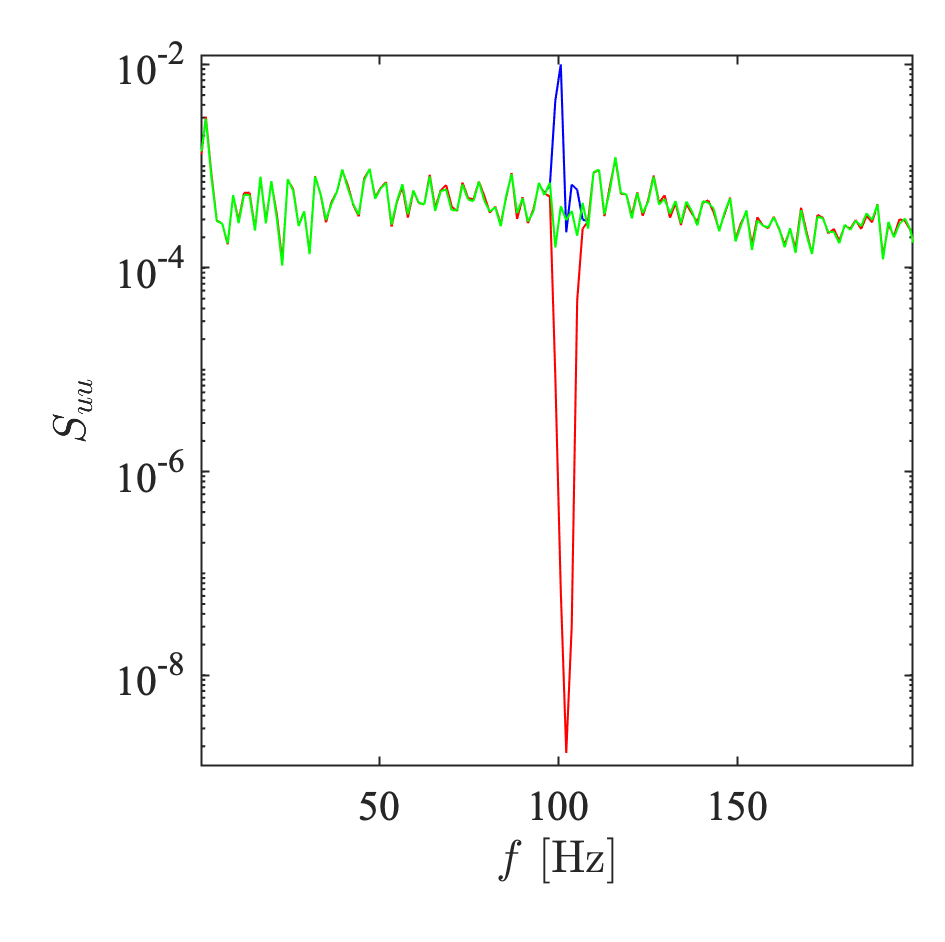}
		\caption{Zoom-in on the filtered frequency range.}
		\label{fig:f32}
	\end{subfigure}
	\caption{Comparison between traditional filtering using {MATLAB}'s \texttt{bandstop} and \scout 's filtering method. 
		The spectra of the original signal (blue), of the {MATLAB}-filtered signal (red), and of the filtered signal using \scout (green) are shown. 
		Also shown (b), is a zoom-in on the filtered frequency range. } \label{fig:Filters}
\end{figure}

The user is subsequently offered the option to filter the detected (or any other) frequency band using a specially-modified Fourier filter.
Unlike typical filters, such as Butterworth or Chebychev, which require a lot of tweaking and tend to over-attenuate the spectrum in the target frequency band yielding a distorted signal in the time domain, the modified Fourier filter is very easy to set and yield a smooth spectrum.
The filter simply attenuates the desired frequency range by replacing the Fourier coefficients at those frequencies with a uniformly-distributed \emph{random noise} that have a user-selected standard deviation.
The addition of random noise acts as an alternative to windowing \cite{kemao2015}, and minimizes ripple effects.
The attenuation magnitude ranges between level 0, where the Fourier coefficients are replaced with random noise with low standard deviation, 
and level 1, where the Fourier coefficients are replaced with random noise whose standard deviation is similar to the detected peak magnitude.
A comparison between a traditional and the proposed filtering technique is presented in Figure~\ref{fig:Filters}, where the spectra of the original signal (blue), 
of the {MATLAB}-filtered signal (red), and of the filtered signal using \scout (green) are shown. 
A zoom-in on the filtered frequency range in Figure~\ref{fig:Filters} (b) clearly shows the over-attenuation when using  {MATLAB}'s \texttt{bandstop} despite setting the steepness and the attenuation to the very low levels of 0.5 and 10, respectively.
%On the other hand, \scout yields a continuous spectrum.
% ============================== FIGURE===============================

Despite windowing and overlapping, spectral edge-effects are sometimes unavoidable.
This issue becomes particularly clear when transforming the signal back to the time domain.
\scout addresses this issue, by offering the user to append points at both ends of the sample record \emph{before} filtering is initiated \cite{mayhew1985}.
These appended points are mirror samples, such that for $K$ appended points on each end, 
the signal becomes 
\[
\mathbf{x} = \{ x_K,\, x_{K-1}, \dotsc, x_{1},\, x_{1},\, x_2, \dotsc, x_{N-1},\, x_N,\, x_{N-1}, \dotsc, x_{N-K-1},\, x_{N-K} \} \,,
\]
where $x_1$ and $x_N$ are the original and first and final sample, respectively.
After time-domain reconstruction, the appended samples are chopped and the sample record is returned to its original length.

In {SCOUT GUI}, the detected peak frequency is registered in the `Signal summary' section.
When applicable, the `Signal statistics' section gets also updated with the new statistics of the filtered signal.  

In {SCOUT Script} the spectrum computation and the filtering process is controlled by the following variables:
\begin{itemize}
	\item \texttt{I.Periodicity.Window}: the window type used for the power spectrum.
	\item \texttt{I.Periodicity.NumberOfBlocks}: the number of blocks to divide the sample record.
	\item \texttt{I.Periodicity.Overlap}: the percent amount of overlap between the blocks.
	\item \texttt{I.Periodicity.DiscreteFFT}: the number of discrete FFT points.
	\item \texttt{I.Periodicity.RemovePeriodic}: a binary selection \texttt{Yes/No} on whether to band-stop filter the signal.
	\item \texttt{I.Periodicity.MinLimit}: the lower frequency of the stopband frequency range.  
	\item \texttt{I.Periodicity.MaxLimit}: the upper frequency of the stopband frequency range.  
	\item \texttt{I.Periodicity.Attenuation}: the filter attenuation level.
	\item \texttt{I.Periodicity.Points}: the number of appended points. The default number is  $0.03 \,N$.
\end{itemize}

% =====================================================================
%                                                             PERIODICITY
% =====================================================================
\subsection{Spectral Analysis}
In this section, we detail the spectral analysis that consists of computing the autospectral density function
and the autocorrelation coefficient, and estimating the integral time scale. 
The integral time scale is relevant for physical insights, and for identifying the number of independent samples, which are necessary for the uncertainty analysis.
The integral time scale is defined as
\begin{equation}
I = \int_0^{\tau_{\max}} \rho(\tau) d\tau\,,
\end{equation}
where $\tau_{\max}$ is the maximum time lag, and $\rho$ is the autocorrelation coefficient,
\begin{equation}
\rho = \frac{R_{xx}}{s^2}\,,
\end{equation}
where $R_{xx}$ is the autocorrelation, and $s$ is the standard deviation.
In {SCOUT}, $R_{xx}$ is computed with an indirect approach from the inverse Fourier transform of the autospectral density function \cite{bendat_random_2011}.
This approach is computationally faster and makes use of block averaging, delivering a smoother autocorrelation distribution.

Executing the analysis in {SCOUT GUI} computes and visualizes the auto-spectral density and the autocorrelation coefficient function.
The integral time scale value is displayed on the right of the autocorrelation coefficient plot alongside the maximum time lag $\tau_{\max}$.
The default $\tau_{\max}$ is set to the nearest point around the third zero-crossing of $\rho$.
The integrated area is colored turquoise.
If the third zero-crossing is not satisfactory, the user can change the integration range manually by shifting the slider next to the autocorrelation coefficient plot.
The final computed integral time scale is registered in the `Signal summary' section.

In {SCOUT Script}, the options for computing the autospectral density function are the same as those listed for the `Periodicity' analysis' in section~\ref{sec:Periodicity} (\texttt{I.Spectral.Window},  \\ \texttt{I.Spectral.NumberOfBlocks}, \texttt{I.Spectral.Overlap}, \texttt{I.Spectral.DiscreteFFT}). 
Two additional options are required for computing the autocorrelation coefficient and the integral time scale:
\begin{itemize}
	\item \texttt{I.Spectral.TimeShift}: the total lag number $L$ ( $L= \tau_{max} / \Delta t$).
	\item \texttt{I.Spectral.AutocorrNumOfBlocks}: the total number of blocks for computing the autocorrelation.
\end{itemize}
	
% =====================================================================
%                                                             SECTION
% =====================================================================
\section{Uncertainty Analysis}
This section describes the uncertainty analysis of individual and of propagated signals.
The analysis yields uncertainty quantification of the mean result and of higher-order central moments up to the fourth.
Due to its ease-of-use and direct interpretability, the propagation of uncertainty is performed using the first-order Taylor series method \cite{tukey1957, helton2003}.
The uncertainty is propagated through a data reduction equation (DRE) of the form
\begin{equation*}
r = r(\text{Signal1}, \text{Signal2}, \dotsc, \text{Signal5})\,,
\end{equation*}
where $r$ is the result, and Signal1, $\dotsc$, Signal5 are the uncertain depend variables.
To keep the analysis simple and approachable, the uncertainty propagation analysis is limited to the most practical and recommended approaches:
\begin{enumerate}
	\item The random uncertainty is computed directly on the result $r$, as recommended \cite{bendat_random_2011}. 
	In other words, individual random uncertainties are not propagated through the DRE, thus bypassing the need to estimate possible correlations among them.
	This, however, requires that all dependent signals are sampled at (or decimated to) the same sample rate and record length.
	\item The systematic uncertainty is only propagated through one accepted functional form of the DRE,
	\begin{equation*}
	r=C(\text{Signal1})^{n1}(\text{Signal2})^{n2}(\text{Signal3})^{n3} \,,
	\end{equation*}
	where $C$ and the $n$'s are user-defined constants. For example, the drag coefficient equation can be expressed as $C_D=2F_D\rho^{-1}V^{-2}A^{-1}=(2\rho^{-1}A^{-1})F_DV^{-2}$, where $F_D$ is the measured force from a balance, $\rho$ is the measured flow density, $V$ is the free stream velocity, and $A$ is the projected area of the model. Assuming a constant flow density and a constant model surface area, this yields
	\begin{itemize}
		\item The result $r=C_D$, which is the variable into which we want to propagate the uncertainty
		\item $C=2\rho^{-1}A^{-1}$, which is constant
		\item $\text{Signal1} = F_D, \quad n1=1$
		\item $\text{Signal2} = V, \quad n2=-2$
	\end{itemize}
	It is worth to note that when the uncertainty of an individual (non-propagated) signal is desired (e.g. only that of Signal1), the user can simply set the DRE constants to unity, i.e. $C=1$ and $n1=1$.
	\item A large sample record is assumed, which yields a coverage factor $t=2$, i.e., an expanded uncertainty $U_r=2\, u_r$, where $u_r$ is the combined uncertainty of the result.
	\item All reported uncertainties are for 95\% confidence level, which is typical for engineering applications.
\end{enumerate}

In {SCOUT}, the random uncertainty of higher-order central moments can be computed using two approaches.
The first one employs simplified equations for the second and fourth central moments,
\begin{subequations}\label{eq:HigherMomentsGaussian}
	\begin{align}
	u_{x^2}=\sqrt{\frac{2}{N_{eff}}}\,,\\
	u_{x^4}=\sqrt{\frac{11}{N_{eff}}}\,,
	\end{align}
\end{subequations}
which assumes a normally-distributed signal.
$N_{eff}$ refers to the effective \emph{statistically independent} number of samples, which is estimated as
\begin{equation*} \label{eq:Neff}
N_{eff}=\left\{
\begin{array}{ll}
\frac{T}{2I}=\frac{N\>\Delta t}{2\>I} \quad \text{if} \quad \Delta t < 2\>I\\
N             \quad \text{if} \quad \Delta t \geq 2\>I
\end{array}
\right.
\end{equation*}

where $T$ is the total sampling time.
For a Gaussian process, the odd moments are identically zero.
Alternatively, the uncertainties of all central moments can be estimated directly with
\begin{equation}\label{eq:DirectEq}
u_{x^m}=\sqrt{\frac{1}{N_{eff}}\frac{\left< x^{2\>m} \right> - \left< x^{m} \right>^2 }{\left< x^{m} \right>^2}}\,,
\end{equation}
where $\left<  \quad \right>$ is the time-averaging operator.
Direct equation~\eqref{eq:DirectEq} is only recommended when the signal(s) is (are) sampled for a sufficiently long time.

Executing the uncertainty analysis in {SCOUT GUI} yields two plots and various outputs:
\begin{itemize}
	\item The time trace of the result $r$ in the center top.
	\item The autocorrelation coefficient distribution of the result in the center bottom. Similar to the `Spectral analysis' tab, the user can adjust the integration range using the slider next to the figure.
	\item The data reduction equation in a readable simplified format in the top right corner.
	\item The random uncertainty of the individual signals as well as of the result.
	\item The systematic uncertainty of the result.
	\item The combined and expanded uncertainty of the result mean $\overline{r}$, which is also used to formally present the mean with its uncertainty band $\overline{r}=r_{mean}\pm U_{\overline{r}}$.
	\item The integral time scale for the result.
	\item Uncertainties of the second, third and fourth central moments using either method.
\end{itemize}

The entire uncertainty analysis in {SCOUT Script} is set with one type 2 configuration file.
In the following, we list the setting options:
\begin{itemize}
	\item \texttt{U.Uncertainty.Coefficient}: the constant $C$ in the data reduction equation.
	\item \texttt{U.Uncertainty.InputSystematic}: a binary selection (\texttt{Yes/No}) on whether to include the systematic uncertainties. 
	\item \texttt{U.Uncertainty.TimeShift}: the total lag number $L$.
	\item \texttt{U.Uncertainty.Exponent1} (and other similar variables): the corresponding signal exponent in the data reduction equation.
	\item \texttt{U.Uncertainty.Systematic1} (and other similar variables): the relative systematic uncertainty of the corresponding signal.
	\item \texttt{U.Uncertainty.AutocorrNumOfBlocks}: the total number of blocks for the autospectral density function to compute the autocorrelation of the result.
	\item \texttt{U.SecondCentral.Normal} (and other similar variables): binary selection (\texttt{Yes/No}) on whether to compute the second order central moment considering normal distribution.
	\item \texttt{U.SecondCentral.Formula} (and other similar variables): binary selection (\texttt{Yes/No}) on whether to compute the second order central moment using a direct equation.
\end{itemize}

%% -- Summary/conclusions/discussion -------------------------------------------

\section{Conclusions} \label{sec:conclusions}
Signal processing and uncertainty quantification and propagation are fundamental fields for the engineering sciences.
However, the wealth of information and the complexity of methods are hindering broad adaptation.
Thus, a compact, interactive, and easy-to-use toolbox provides an attractive alternative to existing expansive packages, which typically require expert knowledge.
The price to pay is the limited offered options for signal processing and for uncertainty analysis, which can be easily remedied through own-code integration.
With its two flavors, \scout offers unique capabilities for both interactive and integrated analysis.
We believe \scout could well be a suitable first choice for data analysis and uncertainty quantification.

\scout builds on years of experience and best practices in processing experimental data.
It covers a range of analyses typically encountered when post-processing measured signals.
These include stationarity analysis, spurious samples detection, normality check, periodicity check, filtering, spectral analysis,
and uncertainty analysis of individual and of multiple propagated signals through a data reduction equation.
Numerical checks show the consistency and validity of the results.

Future developments on \scout include more spectral analysis options, such as wavelet transform, and expanding the uncertainty propagation analysis to include more general functional forms and direct computation of the gradient for the Taylor series expansion method.

%% -- Optional special unnumbered sections -------------------------------------

%\section*{Acknowledgments}

%% -- Bibliography -------------------------------------------------------------
%% - References need to be provided in a .bib BibTeX database.
%% - All references should be made with \cite, \citet, \cite, \citealp etc.
%%   (and never hard-coded). See the FAQ for details.
%% - JSS-specific markup (, , \texttt) should be used in the .bib.
%% - Titles in the .bib should be in title case.
%% - DOIs should be included where available.
\bibliographystyle{plain}
\bibliography{SCOUT-Literature}

\end{document}